\documentclass[aps,twocolumn,superscriptaddress]{revtex4}

\usepackage{epsfig}
\usepackage{color}
\usepackage{graphicx}
\usepackage{dcolumn}
\usepackage{bm}
\usepackage[normalem]{ulem}
\usepackage{textcomp}
\usepackage{soul}
\usepackage[dvipsnames]{xcolor}
\usepackage{amsmath}
\usepackage{amsfonts}
\usepackage{hyperref}
\usepackage{gensymb}
\hypersetup{colorlinks=true,
            linkcolor=blue,
            citecolor=blue,
            urlcolor=orange}

\mathchardef\mhyphen="2D

\makeatother

\begin{document}

\title{
Magnetism and symmetry of superconducting gap in LaFeAsO\\ from dynamical mean-field theory 
}

\author{S. L. Skornyakov}
\affiliation{M. N. Mikheev Institute of Metal Physics of Ural Branch of Russian Academy of Sciences, S. Kovalevskaya Street 18, 620990 Yekaterinburg, Russia}
\affiliation{Ural Federal University, 620002 Yekaterinburg, Russia}

\author{V. I. Anisimov}
\affiliation{M. N. Mikheev Institute of Metal Physics of Ural Branch of Russian Academy of Sciences, S. Kovalevskaya Street 18, 620990 Yekaterinburg, Russia}
\affiliation{Ural Federal University, 620002 Yekaterinburg, Russia}

\author{A. A. Katanin}
\affiliation{Moscow Institute of Physics and Technology, Institutsky lane 9, Dolgoprudny, 141700, Moscow region, Russia}
\affiliation{M. N. Mikheev Institute of Metal Physics of Ural Branch of Russian Academy of Sciences, S. Kovalevskaya Street 18, 620990 Yekaterinburg, Russia}

\date{\today}

\begin{abstract}
By employing a combined method of density functional theory and 
dynamical mean field theory (DFT+DMFT) we investigate the effect 
of electronic correlations on the magnetic and superconducting 
properties of the iron-based parent compound LaFeAsO. We find that 
the static non-local susceptibility $\chi({\bf q})$  
and the dynamical spin structure factor $S({\mathbf q},\omega)$ exhibit a peak 
at the in-plane wave vector ${\mathbf Q}=(\pi,\pi)$, which is strongly 
enhanced upon inclusion of dynamical vertex corrections in the ladder 
approximation, leading to magnetic instability. Considering the 
eigenfunctions of the Bethe-Salpeter equation with the vertex, 
obtained within the second order perturbation theory, as well as 
the ladder approach containing dynamic interaction vertices, in 
agreement with earlier weak-coupling-based studies of LaFeAsO, 
we obtain a close competition between $d$-wave and $s_{\pm}$ order 
parameters, dominating in the second-order and ladder approach, 
respectively. We argue that the dominating $s_{\pm}$ instability in 
the ladder DFT+DMFT approach is related to the reduced degree of 
magnetic frustration by itinerant degrees of freedom due to only 
partially formed local magnetic moments. Our study shows that dynamic 
correlation effects do not change the type of the leading superconducting 
instability in LaFeAsO.
\end{abstract}

\maketitle

\section{INTRODUCTION}

The discovery of the superconductivity of fluorine-doped LaFeAsO 
with the critical temperature $T_c\sim 26$~K initiated a tremendous 
research activity in the field of iron-based superconductors 
(FeSC)~\cite{Kamihara_JACS_2008,Stewart_RMP_2011,Johnston_AdPh_2010,Chubukov_phtd_2015}. 
Since then, extensive efforts have been directed toward the synthesis 
of new FeSC, and theoretical and experimental analysis of the complex 
interplay between structural, electronic, and magnetic properties 
of both the normal and superconducting states of these 
materials~\cite{Dai_RMP_2015,Hosono_physc_2015,Fernandes_Nature_2022}.

An important question in the physics of the superconducting state of 
FeSCs concerns the symmetry of the superconducting order parameter and 
its relation to magnetic properties. There is strong experimental 
evidence for spin-singlet pairing with a sign-changing gap in the main 
families of FeSCs~\cite{spm_exp}. However, identifying the precise 
momentum dependence of the gap is rather challenging and requires 
sophisticated experimental techniques. In the majority of cases, 
results of these experiments are consistent with the so-called $s_{\pm}$ 
scenario, with the isotropic gap changing sign between Fermi surface (FS) 
pockets~\cite{spm_proposal}.

From the theoretical side, it is rather widely accepted that in 
FeSC, spin fluctuations are the most likely candidates to mediate 
superconducting pairing in the presence of strong Coulomb repulsion 
between Fe $3d$ electrons. This is illustrated, for example, by 
the widely used weak coupling spin-fluctuation approach, treating 
the electron interaction in the Fe $3d$ states within the multi-band 
Hubbard Hamiltonian and employing the random-phase 
approximation~\cite{Graser_NJP_2009,Kemper_NJP_2010,Maier_prb_2011} 
or functional renormalization group approach \cite{fRG} to calculate 
the pairing vertex. Alternative pairing concepts include the orbital 
fluctuation approach, treating charge and spin fluctuations on an 
equal footing, and strong coupling theories assuming nearly localized 
magnetism of FeSC~\cite{Misawa_ncomm_2014,sc_orb_fluct,sc_strng_cpl}.

Interestingly, there is no universal conclusion on the pairing symmetry 
of FeSC, even within the same pairing theory. For example, numerical 
results for tight-binding models derived from the electronic structure 
obtained by density functional theory (DFT) in combination with the 
spin-fluctuation approach predict competing $s_{\pm}$ and $d-$ wave 
superconductivity of LaFeAsO depending on the values of the Coulomb 
interaction parameters~\cite{Graser_NJP_2009}. For a structurally 
related compound FeS which exhibits a configuration of the FS similar 
to that of LaFeAsO (with both electron and hole pockets) 
the same approach yields a crossover from $d_{x^2-y^2}$ to $s_{\pm}$ 
symmetry of the order parameter under pressure~\cite{Shimizu_prl_2018}.

The DFT-based weak coupling pairing theories are challenged by the 
observation that electron interactions in the partially filled Fe~$3d$ 
shells of FeSC are actually not weak as evidenced e.g. by renormalizations 
of the band mass compared to DFT values~\cite{Si_NJP_2009,Hund}. 
In fact, FeSCs belong to the class of so-called Hund 
metals where correlated electron behavior originates not only from 
Coulomb repulsion but is largely mediated by Hund's rule coupling~\cite{hund_metals}. 
This raises a question of the adequacy of tight-binding models derived 
from DFT and highlights the role of correlation effects in accurately 
describing the electronic properties of FeSC. For instance, 
state-of-the-art approaches combining the dynamical mean-field theory 
of correlated electrons and band methods (DFT+DMFT) \cite{KotliarRev,DMFT1} 
have demonstrated a systematic improvement in predicting the electronic 
structure and the lattice and magnetic properties of FeSC over DFT 
estimates~\cite{Backes_NJP_2014,Watson_PRB_2017,Skornyakov_PRB_2009}.
However, most of the DFT+DMFT calculations reported in the literature 
focus primarily on the normal state of FeSC, with relatively few 
studies addressing the superconducting properties of iron-based 
materials within this method~\cite{DFT_DMFT_sc}. We note that in 
these calculations the particle-hole irreducible vertex is taken 
into account in a static approximation neglecting its frequency 
dependence, which may significantly alter numerical results for 
correlated materials.

Here, we investigate the effect of dynamic vertex corrections 
by examining the interplay between magnetic properties and the 
superconducting gap structure of the prototypical iron pnictide parent 
compound LaFeAsO. To this end, we employ two complementary approaches 
within DFT+DMFT: a second-order perturbation theory that incorporates 
the electronic self-energy as obtained by DFT+DMFT, and a diagrammatic 
ladder-type vertex approximation which is a standard tool for studying 
two-particle properties within DFT+DMFT~\cite{KotliarRev,OurRev}. Due to 
the recent methodological developments, that made it possible 
to compute the full frequency dependence of particle-hole irreducible 
vertices, the latter approach has been successfully applied to a 
number of iron- and chromium-based magnetic materials~\cite{OurJq,Fe2C,CrO2,CrTe2,Fe3GeTe2}.  
Using these methods, we study the possibility of superconducting pairing 
in LaFeAsO in the presence of both self-energy and {\it dynamic vertex corrections}, 
arising from the strong electronic correlations inherent to this compound.

The paper is organized as follows. In Sec.~\ref{tech_det} we provide 
technical information on our DFT+DMFT calculations and briefly describe 
the method we use to compute the superconducting gap. The DFT+DMFT 
results for the spectral and magnetic properties, including the 
electronic structure and nonuniform spin susceptibility, are presented 
in Sec.~\ref{fs_suscep}. Preferable symmetries of superconducting 
pairing are discussed in Sec.~\ref{sec_gap}. Finally, the results are 
summarized in Sec.~\ref{sec_concl}

\section{Method and Computational details}
\label{tech_det}
In this work, the electronic properties of LaFeAsO are investigated 
by the DFT+DMFT method \cite{KotliarRev,DMFT1} implemented within the 
pseudopotential approach and generalized gradient approximation in 
DFT~\cite{GGA,sc_dmft}. The calculations were performed for the room 
temperature tetragonal crystal structure of LaFeAsO (space group 
$P4/nmm$) with the unit cell parameters $a=b=4.0322$~$\mathrm\AA$, 
$c=8.7364$~$\mathrm\AA$ and the internal coordinate of As 
$z_{\mathrm{As}}=0.6508$ taken from Ref.~\cite{Qureshi_prb_2010}.
The DMFT equations are solved with the effective Wannier function–based 
Hamiltonian $\hat H_{\mathrm{DFT}}({\bf k})$ obtained by the projection 
procedure~\cite{Wannier2}. To construct $\hat H_{\mathrm{DFT}}({\bf k})$ 
we use an energy window spanning the complex of bands originating from 
the Fe-$3d$, O-$2p$ and As-$4p$ orbitals. For the Coulomb repulsion, 
we consider a density-density Hamiltonian
\begin{equation}
H_{\rm int}=\frac{1}{2}\sum\limits_{i,mm',\sigma\sigma'} U^{mm^\prime}_{\sigma\sigma^\prime}
{n}_{im\sigma} {n}_{im^\prime\sigma^\prime},
\label{Hint}
\end{equation}  
where $n_{im\sigma}=c^+_{im\sigma} c_{im\sigma}$ ($c^+_{im\sigma}$ 
and $c_{im\sigma}$ are the electron creation and destruction operators, 
the indices $i$ and $m$,$m'$ numerate, respectively, the unit cells and 
$d$-orbitals of iron atoms in the unit cell); the matrix $U^{mm'}_{\sigma\sigma'}$ 
contains only intra-atom interactions and parameterized 
by the Slater integrals $F^i$. The DMFT impurity problem was solved 
by the continuous-time quantum Monte-Carlo method~\cite{ctqmc}. 
In these calculations, we take Hubbard $U=F^0=3.5$~eV and Hund 
exchange $J=(F^2+F^4)/14=0.85$~eV close to those used in previous 
DFT+DMFT studies of spectral and magnetic properties of LaFeAsO and 
other iron pnictides and chalcogenides~\cite{Anisimov_JPCM_2009,Aichhorn_PRB_2009,Skornyakov_PRB_2019,Skornyakov_PRB_2021}.

We determine the pairing vertex $V_{{\bf k}{\bf k}^{\prime}}$ (${\bf k}$, ${\bf k}^{\prime}$ 
being the corresponding momenta at FS) in two different ways: we 
consider the second-order perturbation theory ~\cite{Belozerov_PRB_2020}, as well 
as the ladder DMFT approach \cite{OurJq}. 

The pairing vertex in the second order perturbation theory was obtained 
in Ref. \cite{Belozerov_PRB_2020} (see also Appendix for the density-density form). In the ladder approach, we obtain the 
{following expression for the multi-orbital} interaction in the singlet 
pairing channel (cf. Ref. \cite{Kitatani} for the single-band case)
\begin{equation}
V^{mm'}_{\mathbf{kk}'}=\frac{1}{4}\left(3F^{s,mm'}_{\mathbf{k}-\mathbf{k}'}+3F^{s,mm'}_{\mathbf{k}+
\mathbf{k}'}-F^{c,mm'}_{\mathbf{k}-\mathbf{k}'}-F^{c,mm'}_{\mathbf{k}+\mathbf{k}'}\right)
\end{equation}
where we have omitted all local contributions, since we are interested 
in unconventional pairing (we have verified that they produce negligible 
contributions),  
$F^{s(c)}_{\mathbf q}=[\delta_{\nu\nu''}I-\Gamma^{s(c)}_{{\rm ir},\nu\nu''}\chi^{0}_{\nu''\mathbf {q}}]^{-1}\Gamma^{s(c)}_{{\rm ir},\nu''\nu}$ is the interaction vertex 
in the spin (charge) channel (continued to the frequency $\nu=0$), $I$ is 
the unity matrix in orbital indices; matrix multiplications and inversions with respect to frequency and orbital indices are assumed, $\Gamma^{s(c),mm'}_{{\rm ir},\nu\nu'}$ is the respective local 
irreducible vertex, extracted from the DMFT impurity problem, and 
$\chi^{0,mm'}_{\nu\mathbf {q}}=-k_{\mathrm B}T\sum_{\mathbf{k}} G^{mm'}(\mathbf{k},\nu)G^{m'm}(\mathbf{k+q},\nu)$ 
is the bare frequency-resolved static susceptibility, 
$G^{mm^{\prime}}({\bf k},i\nu_n)=[(i\nu_{n}+E_{\mathrm F})I +M_{\rm DC}- H_{\mathrm{DFT}}({\bf k})-\Sigma(i\nu_{n})]^{-1}_{mm^{\prime}}$ is the DFT+DMFT Green's function 
matrix, $M_{\rm DC}$ is the double counting correction. 

To identify possible symmetries of superconducting pairing, we compute 
eigenvalues $\lambda_{\alpha}$ and eigenfunctions $f_{{\bf k},\alpha}$ 
of the multiband Bethe-Salpeter equation 
\begin{equation}
  \lambda f_{{\bf k},\alpha}
  =
  -k_{\mathrm B}T\sum_{\substack{{\bf k}^{\prime} \\ i\nu_{n},\beta}}
  {V}_{{\bf k}{\bf k}^{\prime}}^{\alpha\beta}
  f_{{\bf k}^{\prime},\beta}
  G_{{\bf k}^{\prime}\beta}(i\nu_{n})
  G_{-{\bf k}^{\prime}\beta}(-i\nu_{n}),
 \label{BSE}
\end{equation}
where $G_{{\bf k}\alpha}(i\nu_{n})$ is the Green function of band $\alpha$ 
obtained by diagonalization of the DFT+DMFT Green function matrix (with the 
self-energy approximated by $\mathrm{Re}\Sigma(0)$), $\nu_{n}$ is the 
fermionic Matsubara frequency, and $E_{\mathrm F}$ is the Fermi energy. 
To pass to the band indices in the interaction, we use the matrices 
$S^{m\alpha}_{\mathbf k}$, obtained as eigenvectors of the effective Hamiltonian 
$H(\mathbf{k})=H_{\mathrm{DFT}}({\bf k})+{\rm Re}\Sigma(0)-M_{\rm DC}$, 
cf. Refs. ~\cite{Belozerov_PRB_2020,Graser_NJP_2009}, and gauged to preserve 
the continuity of their phases. Within this approach the leading eigenvalue 
corresponds to the preferred pairing symmetry and the respective eigenfunction 
$f_{{\bf k},\alpha}$ determines the momentum dependence of the superconducting 
gap. 

\begin{figure}[t]
\centering
\includegraphics[width=0.45\textwidth,clip=true,angle=0]{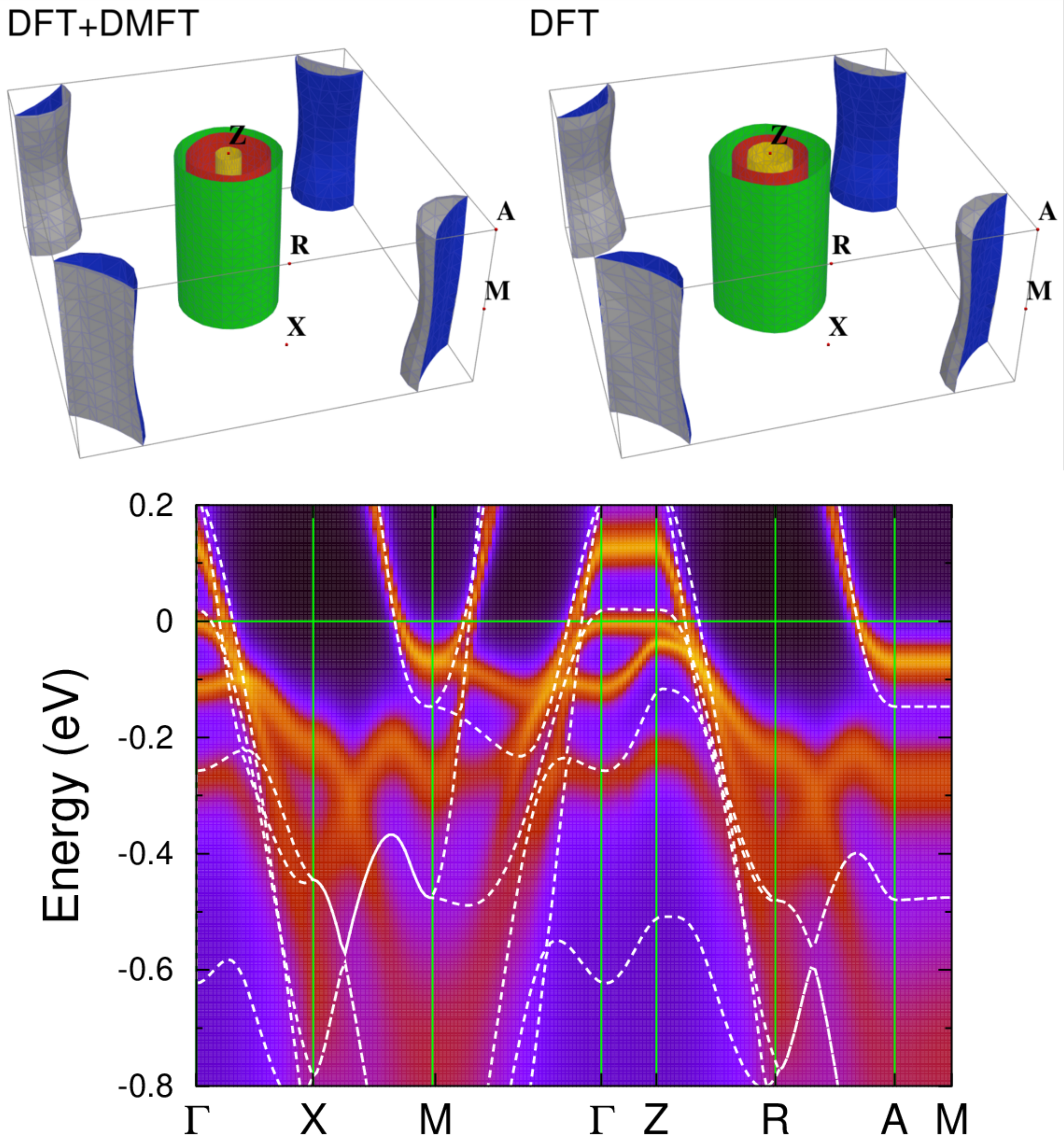}
\caption{
(Color online). Top: Fermi surface of LaFeAsO as calculated by DFT and 
DFT+DMFT. Bottom: momentum-resolved spectral function $A({\bf k},\nu)$ 
of LaFeAsO obtained by DFT+DMFT (contours) in comparison with the band 
structure computed by DFT (dashed lines). The DFT+DMFT calculations 
correspond to the temperature $T=290$~K ($\beta=40$~eV$^{-1}$). The Fermi 
energy is set to $0$~eV.
}
\label{fig_1}
\end{figure}

\section{Results}
\subsection{Fermi surface and magnetic properties}
\label{fs_suscep}
As a starting point, we evaluate the effect of Coulomb correlations 
on the electronic structure of LaFeAsO. In Fig.~\ref{fig_1} 
we compare the band structure and the Fermi surface (FS) of LaFeAsO 
computed by DFT and DFT+DMFT. Due to the finite quasiparticle lifetime 
within DFT+DMFT the band structure is represented by the momentum-resolved 
spectral function $A({\bf k},\nu)$,
\begin{equation}
A({\bf k},\nu)=-\frac{1}{\pi}\sum_{i} \mathrm{Im} G^{ii}({\mathbf k},\nu),
\end{equation}
obtained by analytic continuation of $\hat\Sigma(i\nu_{n})$ 
using Pad\'e approximants~\cite{Pade}, and the FS is determined 
by the condition $H(\mathbf{k})=E_{\mathrm F}$.

\begin{figure}[t]
\centering
\includegraphics[width=0.58\textwidth,clip=true,angle=-90]{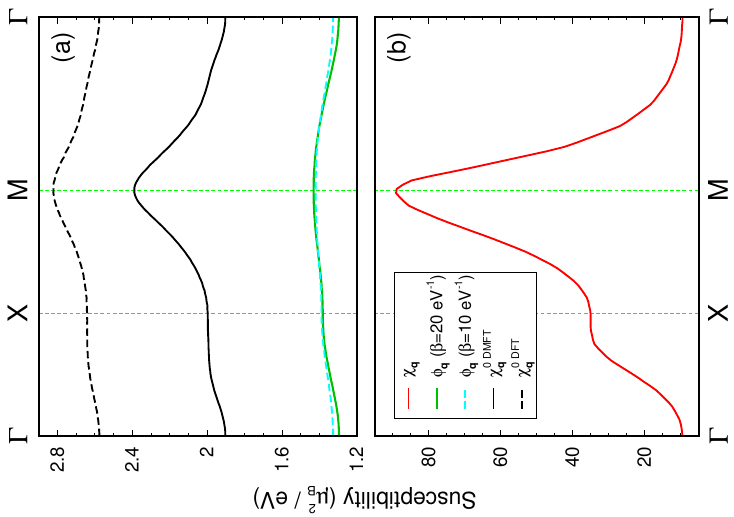}
\caption{
(Color online). (a) Momentum dependence of the static particle-hole bubble 
$\chi^{0}_{\bf q}$ as obtained by DFT (black dashed line) and DFT+DMFT 
(black solid line) at $T=290$~K ($\beta=40$~eV$^{-1}$), compared to 
that of the static irreducible susceptibility $\phi({\bf q})$ in the ladder 
DFT+DMFT approach at $\beta=10$~eV$^{-1}$ (green line) and 
$\beta=20$~eV$^{-1}$ (cyan dashed line). (b) Momentum dependence of 
the nonlocal spin susceptibility in the ladder DFT+DMFT approach at 
$\beta=10$~ eV$^{-1}$.
}
\label{fig_2}
\end{figure}

We observe that within DFT the low-energy band structure of LaFeAsO at 
the $\Gamma$ point features three parabolic hole-like bands crossing 
the Fermi level in the first third of the $\Gamma$-M and $\Gamma$-X 
directions. These bands give rise to three concentric FS cylinders 
centered at the $\Gamma$-Z direction. In addition, there are two elliptical 
intersecting FS sheets centered at the M-A direction originating from 
two electron-like parabolic bands with a minimum at the M point. 

Upon inclusion of electronic correlations, the low-energy electronic structure 
of LaFeAsO undergoes a transformation compared to the result obtained by DFT, 
in agreement with previous DFT+DMFT calculations of LaFeAsO and other parent 
compounds of FeSC~\cite{Aichhorn_PRB_2009,Skornyakov_PRB_2019,Skornyakov_PRB_2021}. 
Specifically, we observe that within DFT+DMFT the radius of the hole-like FS 
cylinders centered at the $\Gamma$-Z direction is significantly reduced 
compared to DFT whereas the size of the electron-like FS at the M-A line is 
almost unaffected by electronic correlations. As a result, the FS computed 
within DFT+DMFT exhibits enhanced nesting with the wave vectors 
$\mathbf{Q}=(\pi/a,\pi/a,q_z)$ connecting electron and hole pockets and 
promoting antiferromagnetic spin fluctuations. 

To further elucidate the character of magnetic fluctuations in the 
paramagnetic state of LaFeAsO we compute the bubble static spin 
susceptibility 
$\chi^0({\bf q})=\sum_{mm',\nu_n} \chi^{0,mm'}_{\nu,\mathbf{q}}$, 
as well as the static orbital summed irreducible susceptibility 
$\phi^{mm'}(\mathbf q)=\sum_{\nu_n,m''} \gamma^{mm''}_{\nu,{\mathbf q}} \chi^{0,m''m'}_{\nu,\mathbf{q}}$, 
which accounts for the vertex corrections $\gamma_{\nu,{\mathbf q}}$, 
and the respective orbital summed nonlocal susceptibility 
$\chi^{mm'}(\mathbf q)=\phi(\mathbf q)[1-U^s \phi(\mathbf q)]^{-1}$ 
in the ladder approximation \cite{OurJq} ($U^s=U_{\uparrow \downarrow}-U_{\uparrow\uparrow}$ 
is the interaction in the spin channel).
Our result for $\chi^0({\bf q})$ along the $\Gamma$XM$\Gamma$ path in the 
Brillouin zone, as obtained by DFT and DFT+DMFT approaches, is shown in 
Fig.~\ref{fig_2} (a). We observe that both methods yield a similar shape 
of $\chi^0({\bf q})$ with pronounced maximum at the M point, consistent 
with the ${\mathbf q}_M=(\pi,\pi,0)/a$ nesting vector of the FS. 
We also note that upon taking into account the self-energy correction in 
the DFT+DMFT approach, the overall magnitude of $\chi^0({\bf q})$ drops by 
$\sim 30$\% while the peak at the M point becomes sharper. Importantly, 
the bare bubble has a substantial contribution at ${\mathbf q}_{X}=(\pi/a,0,0)$. 
The vertex corrections, represented by the difference of $\chi^0({\mathbf q})$ 
and $\phi({\mathbf q})$ further suppress the irreducible susceptibility 
and make it even flatter with respect to the momentum ${\mathbf q}$. 
However, the full (reducible) susceptibility at $\beta=10$~eV$^{-1}$, 
shown in  Fig.~\ref{fig_2} (b) for $q_z=0$, has a strong peak at ${\bf q}_M$. 
The obtained bubble, irreducible and reducible magnetic susceptibilities
only weakly depend on $q_z$. These results demonstrate that the magnetic 
instability of LaFeAsO is indeed promoted by interaction effects and occurs 
at the in-plane wave vector ${\bf Q}$, implying two-dimensional magnetic 
interactions in agreement with experimental and theoretical results for 
structurally related pnictides and chalcogenides.

\begin{figure}[t]
\centering
\includegraphics[width=0.48\textwidth,angle=0]{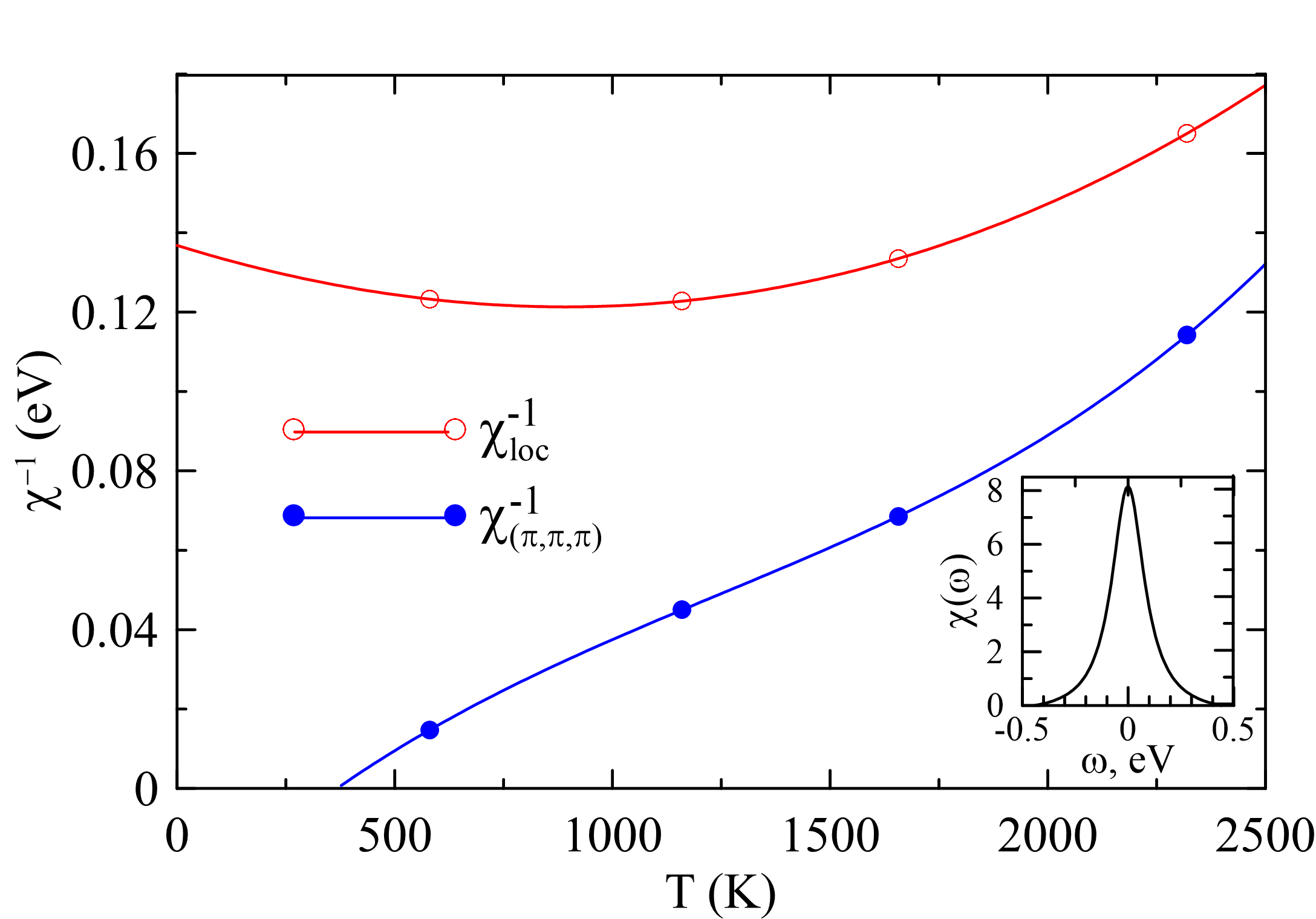}
\caption{
(Color online). Temperature dependence of the inverse local (open circles) 
and staggered (characterized by the wave vector $(\pi/a,\pi/a,\pi/c)$, 
filled circles) susceptibilities in the ladder DFT+DMFT approach. 
The inset shows the dynamic local susceptibility at the real frequency 
axis at $\beta=10$~eV$^{-1}$.}
\label{fig_3}
\end{figure}

The temperature dependence of the local and staggered susceptibilities 
in the ladder DFT+DMFT approach is shown in Fig.~\ref{fig_3}. One can 
see that the inverse local susceptibility shows linear-like behavior 
only at sufficiently high temperatures, signaling that the local moments 
are only partly formed. This is also confirmed by the frequency dependence 
of the local spin susceptibility on the real frequency axis, showing a  
peak which half width at half height $\simeq k_{\mathrm B}T$. The dominating 
spin susceptibility, related to the wave vector $(\pi/a,\pi/a,\pi/c)$ 
diverges at the temperature $T^{\rm DMFT}_N\simeq 375$~K. In view of 
the mean field nature of the DMFT approach, this temperature characterizes 
the formation of a strong short-range magnetic order.

\begin{figure}[t]
\centering
\includegraphics[width=0.45\textwidth,angle=0]{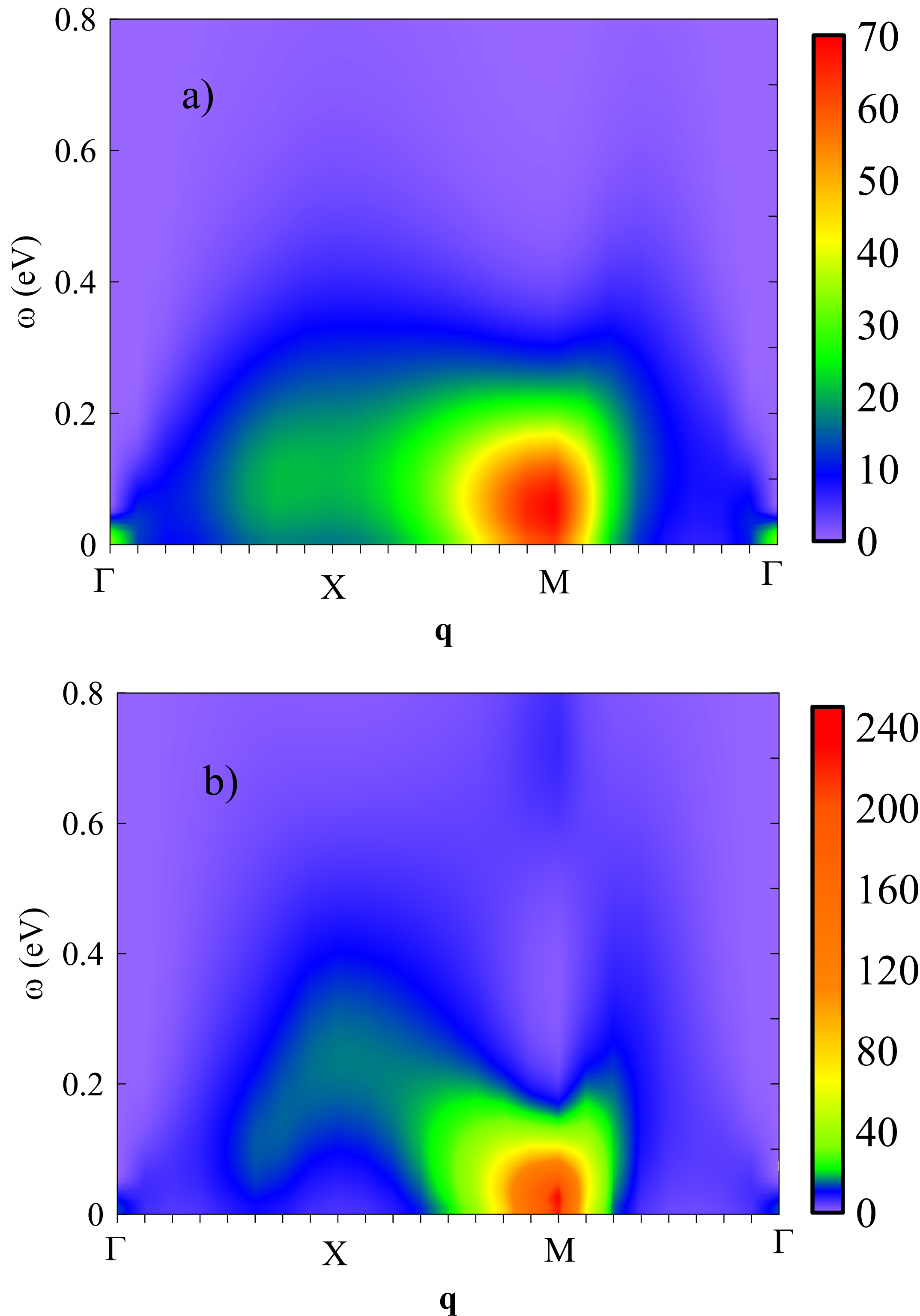}
\caption{
(Color online). Dynamical spin structure factor $S({\mathbf q},\omega)$ 
along high-symmetry directions in the ladder DFT+DMFT approach at 
(a) $\beta=10$~eV$^{-1}$ and (b) $\beta=20$~$eV^{-1}$.
}
\label{fig_4}
\end{figure}

In Fig. \ref{fig_4} we show the dynamical spin structure factor 
$S({\mathbf q},\omega)=(2/\pi)\mathrm {Im}\chi({\mathbf q},\omega)/(1-\exp(-\omega/T))$ 
obtained in the ladder DFT+DMFT approach. One can see that, in accordance 
with the static susceptibilities of Fig. \ref{fig_2}, the maximum of 
$S({\mathbf q},\omega)$ is obtained near $\mathbf{q}=\mathbf{q}_M$, with 
the ``satellite" branch of magnetic excitations falling from the point 
${\mathbf q_X}$ in the $X-\Gamma$ direction, and becoming progressively 
more pronounced with the reduction of the temperature. Although the overall 
shape of $S({\mathbf q},\omega)$ is similar to that obtained previously 
within the static approximation \cite{DFT_DMFT_sc} (up to the interchange 
of the points ${\mathbf q}_X$ and ${\mathbf q}_M$, related to passing from 
the two iron atoms to the single-atom unit cell), we note that the 
considered dynamic approximation of the vertex yields much larger values 
of dynamical spin structure factor, corresponding to strengthening of 
magnetic fluctuations.

\begin{figure}[t]
\centering
\includegraphics[width=0.48\textwidth,angle=0]{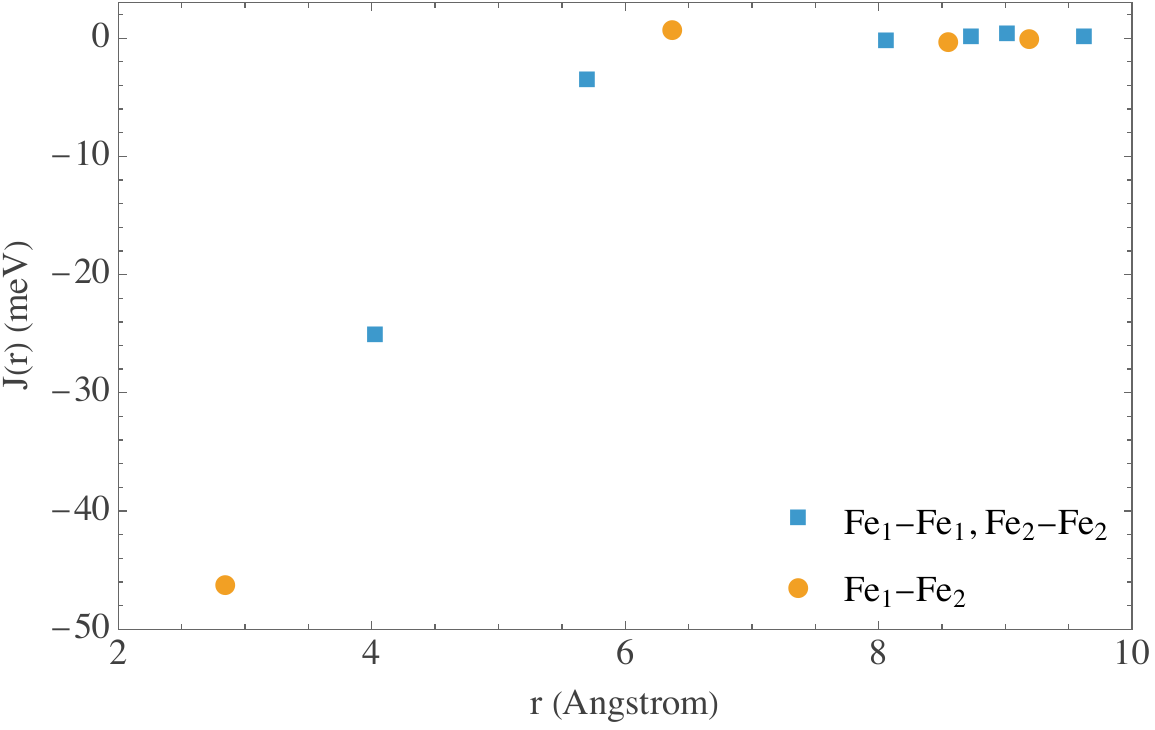}
\caption{
(Color online). Radial dependence of exchange interactions {in LaFeAsO
as obtained by} the ladder DFT+DMFT approach at $\beta=10$~eV$^{-1}$. 
The yellow circles (blue squares) correspond to the interactions between 
Fe$_1$-Fe$_2$ (Fe$_1$-Fe$_1$, Fe$_2$-Fe$_2$) atoms of different (equal) 
sublattices.
}
\label{fig_5}
\end{figure}

Although the local magnetic moments are only partly formed, one can 
estimate the effective momentum-dependent exchange interactions from 
the momentum dependence of the inverse magnetic susceptibility 
$J_{\mathbf q}=\chi_{\rm loc}^{-1}-\chi_{\mathbf q}^{-1}$, where matrix 
inversions with respect to Fe$_{1,2}$ atoms of the unit cell are 
assumed~\cite{OurJq,Fe2C,CrO2,CrTe2,Fe3GeTe2}. This method does not rely 
on the assumption of well-formed local magnetic moments. The Fourier 
transform of $J_{\mathbf q}$, representing the radial dependence of 
exchange interactions, is shown in Fig.~\ref{fig_5}. One can see that 
there is a close competition of the nearest $J_1$ and next-nearest 
neighbor magnetic exchange $J_2$, which are both negative (i.e. antiferromagnetic) 
and sufficiently close  to the frustration point $J_2=J_1/2$ (this is 
also reflected in the flatness of the irreducible susceptibility in 
Fig.~\ref{fig_2} (a)). Proximity of the ratio $J_2/J_1$ 
to 1/2 is however accompanied by the predominant magnetic fluctuations 
at the specific wave vector ${\mathbf q}={\mathbf q}_M$ (see Figs. \ref{fig_2} (b) and 
\ref{fig_4}), which implies that the magnetic degrees of freedom of 
LaFeAsO are not fully described by the effective Hisenberg model. Thus, 
in the present case the system avoids frustration due to itinerant 
degrees of freedom.

\subsection{Superconducting pairing}
\label{sec_gap}

\begin{figure*}[t]
\centering
\includegraphics[width=0.99\textwidth,clip=true,angle=0]{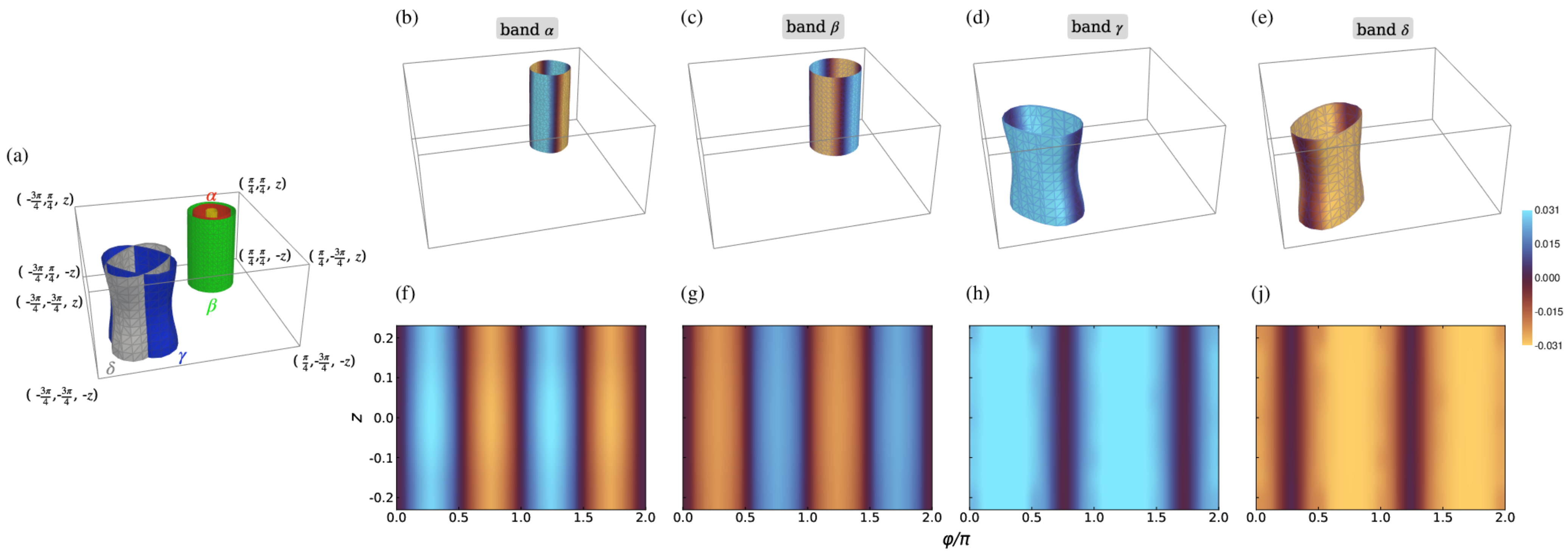}
\caption{
(Color online). (a) Unit cell of the reciprocal space and DFT+DMFT 
Fermi surface of LaFeAsO ($T=290$~K) used for solving the Bethe-Salpeter 
equation. The letters $\alpha$, $\beta$, $\gamma$, and $\delta$ label 
sheets of the FS originating from different bands. [(b)-(e)] Eigenfunction 
of the Bethe-Salpeter equation for singlet pairing as obtained within 
second-order perturbation theory, projected onto FS sheets $\alpha$-$\delta$. 
[(f)-(j)] Same as [(b)-(e)] presented in spherical coordinates $(z,\phi)$. 
The azimuthal angle $\phi=0$ and coordinate $z=0$ correspond to the 
positive direction of $k_{x}$ and the $\Gamma X M$ plane, respectively.
}
\label{fig_6}
\end{figure*}

Next, we investigate the leading pairing instability and estimate 
the symmetry of the superconducting gap of LaFeAsO by computing 
eigenfunctions and eigenvalues of the multiband Bethe-Salpeter 
equation Eq.~\ref{BSE}. To solve Eq.~\ref{BSE}, we shift the 
tetragonal Brillouin zone by a vector $(-\frac{3\pi}{4},-\frac{3\pi}{4},0)$ 
as shown in Fig.~\ref{fig_6}~(a) and consider the four most relevant 
FS sheets linked by the nesting vectors ${\mathbf Q}$: two outermost 
hole-like FS cylinders centered at $\Gamma$-Z and two electronic-like 
FS at the M-A direction. In Fig.~\ref{fig_6}~(a) these sheets are 
labeled by letters $\alpha$, $\beta$, $\gamma$, and $\delta$, respectively. 
This transformation allows for a continuous parametrization of each 
FS in cylindrical coordinates. As the pairing occurs in close vicinity 
of the FS we assume that for each Cartesian $z$ and azimuthal angle 
$\phi$ the gap functions in Eq.~\ref{BSE} are equal to their values 
at the FS (i.e. we neglect their radial dependence in cylindrical 
coordinates, $f_{{\bf k}, \alpha}\to f_{\alpha}(z,\phi)$).

In the second-order perturbation theory for the pairing interaction, 
we find that the leading eigenvalues for the singlet and triplet 
pairing channels, computed within the second-order perturbation theory, 
are $\lambda_s=0.95$ and $\lambda_t=0.90$, respectively, closely 
resembling those obtained from the same approach for 
$\varepsilon$-iron~\cite{Belozerov_PRB_2020}. Although $\lambda_s$ 
exceeds $\lambda_t$ only by about $\sim$5\% the dominance of 
spin-singlet pairing is consistent with experimental observations. 
Furthermore, we note that the presence of competing singlet and 
triplet pairing channels is not unique and has been discussed previously, 
e.g. for Sr$_2$RuO$_4$~\cite{sr2ruo4_ref1,sr2ruo4_ref2}.

The momentum dependence of the spin-singlet gap function in the second 
order perturbation theory is presented in Fig.~\ref{fig_6} (b)-(j) as 
a projection onto the respective FS sheets ({we have verified that the 
shape of both the FS and the gap functions only weakly depends on 
temperature and reaches saturation already at $\beta=40$~eV$^{-1}$}). 
For clarity, the same functions are also displayed in Fig.~\ref{fig_6} 
using cylindrical coordinates $(z,\phi)$. We observe that the singlet 
gap functions exhibit rather weak variation along $k_{z}$ and show 
oscillatory behavior with comparable amplitudes at different parts of 
the FS. Most notably, the sign of the gap varies with $\phi$ at the 
hole-like FS, with lines of nodes located at $\phi=\frac{\pi}{2}n$ ($n\in\mathbb{Z}$). 
In contrast, sign variations on the electronic-like FS are less pronounced. 
Namely, the FS sheet $\gamma$ features a non-negative gap with nodes 
at $\phi=\frac{3\pi}{4}+\pi n$ while sheet $\delta$ shows a non-positive 
gap  with nodes at $\phi=\frac{\pi}{4}n + \pi n$ ($n\in\mathbb{Z}$). 
This sign structure in the ${\bf k}$-space is consistent with the 
$d$-wave pairing symmetry of the superconducting order parameter. 
This result is stable upon switching the 
interaction term to the density-density form (see the Appendix for details).
We also note that the predominance of $d$-wave pairing obtained in our 
calculations aligns with competing $s$ and $d$ pairing channels reported 
in spin-fluctuation theory calculations~\cite{Graser_NJP_2009}.

\begin{figure}[b]
\centering
\includegraphics[width=0.49\textwidth,clip=true,angle=0]{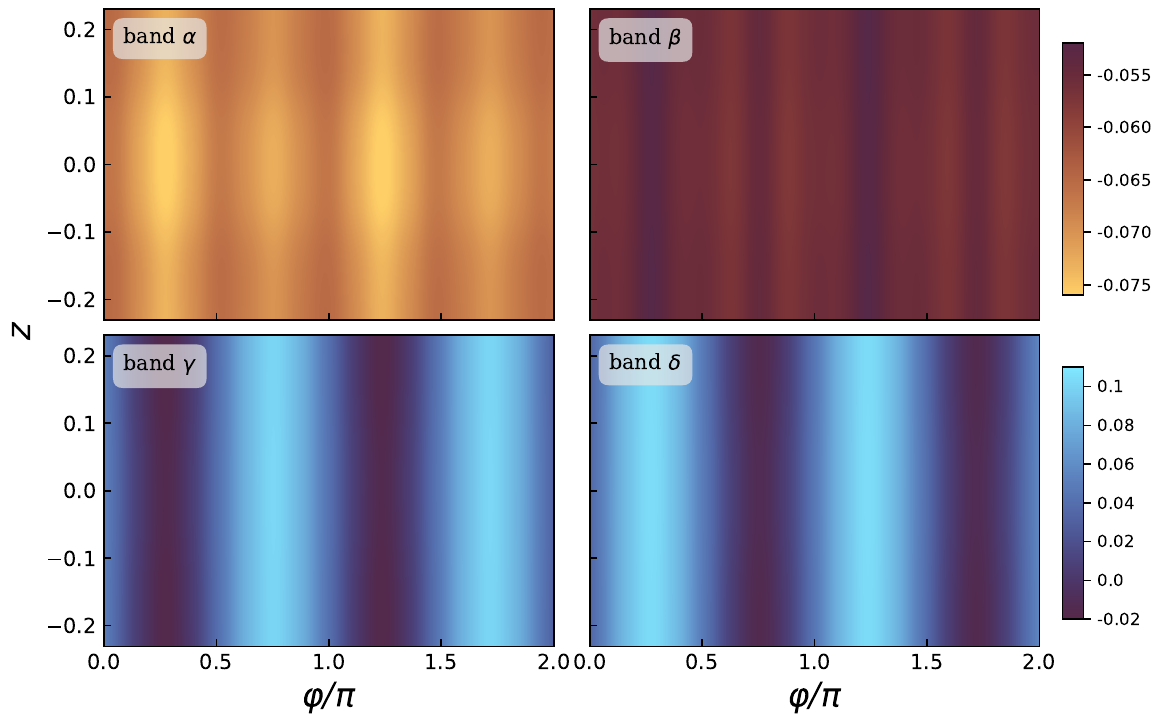}
\caption{
(Color online). Superconducting gap of LaFeAsO shown in cylindrical 
coordinates $(z,\phi)$ as obtained within the ladder DMFT approach. 
The letters $\alpha$, $\beta$, $\gamma$, and $\delta$ refer to same 
FS sheets as those in Fig.~\ref{fig_6}.}
\label{fig_7}
\end{figure}

To gain insight into the role of vertex corrections, we consider the 
results for the obtained gap function in the ladder approximation at 
$\beta=10$~eV$^{-1}$ (see Fig.~\ref{fig_7}). We have verified that 
qualitatively similar result is also obtained at $\beta=20$~eV$^{-1}$. 
In the ladder approach, we 
obtain a dominant pairing of $s_{\pm}$ symmetry with a different 
sign of the gap at $\alpha,\beta$ and $\gamma,\delta$ sheets, similar 
to previous results from the RPA \cite{Graser_NJP_2009} and fRG 
approach~\cite{fRG}. The gap at $\alpha$ and $\beta$ sheets depends 
only weakly on momentum, while in $\beta$ and $\gamma$ sheets, it is 
maximal along the respective diagonal ($\phi=3\pi/4+\pi n$ for $\gamma$ 
sheet and $\phi=\pi/4+\pi n$ for $\delta$ sheet) and minimal along 
the other diagonal. The $d$-wave symmetry of the gap function is 
obtained in this case as the subleading instability in the singlet 
channel. The triplet pairing channel is substantially weakened in 
the ladder DFT+DMFT approach, and its eigenvalue remains an order of 
magnitude smaller than that in the singlet channel.

The competition of the $s_{\pm}$-wave and $d$-wave instabilities can 
be attributed to the competition of magnetic fluctuations with wave 
vectors close to ${\mathbf Q}$ and ${\mathbf Q}_X=(\pi,0,q_z)$. 
Indeed, ${\mathbf q}={\mathbf Q}$ fluctuations promote the interband 
nesting therefore enhance both $s_{\pm}$-wave and, to a lesser 
degree, $d$-wave instabilities. However, the $d$-wave instability is 
additionally enhanced by magnetic fluctuations with wave vectors close 
to ${\mathbf Q}_X$. In the case of second order perturbation theory, 
these fluctuations are comparable to those with ${\mathbf q}={\mathbf Q}$ 
(see Fig.~\ref{fig_2}). The former fluctuations are, however, 
suppressed within the ladder approximation, which yields the $s_{\pm}$ 
superconducting instability.

The abovementioned suppression is closely related to the degree of 
magnetic frustration, discussed in Sect. \ref{fs_suscep}. In the 
second-order perturbation theory, the degree of frustration is large, 
which is directly translated to the exchange interactions of Fig.~\ref{fig_5}. 
Therefore, in this approximation, the system prefers the $d$-wave 
order parameter; the triplet instability is also sufficiently close. 
However, in the ladder DFT+DMFT approach, due to the increased role 
of itinerant degrees of freedom, the frustration is weakened, and the 
$s_{\pm}$ order parameter dominates.

\section{CONCLUSIONS}
\label{sec_concl}

In summary, we have studied the impact of electronic and magnetic 
properties on the symmetry of the superconducting order parameter 
in LaFeAsO using the DFT+DMFT method. We find that the non-local 
particle-hole polarization bubble, the full susceptibility, and 
the dynamical spin structure factor obtained within the ladder 
DFT+DMFT approach, exhibit a peak at ${\mathbf Q}=(\pi/a,\pi/a,q_z)$. 
In the ladder approach, the peak of susceptibility at the wave 
vector $\mathbf{Q}$ is much more pronounced than that at 
$\mathbf{Q}_X=(\pi/a,0,q_z)$. We argue that this behavior physically 
corresponds to a reduced degree of magnetic frustration due to 
itinerant degrees of freedom. Therefore, dynamic correlations in 
LaFeAsO are found to be sufficiently weak to preserve the shape of 
$\chi(\mathbf q)$ comparable to results of weak-coupling approaches.

The obtained momentum dependence of magnetic susceptibility has its 
impact on the symmetry of the gap function. While in the second order 
perturbation theory we find the $d$ -wave order parameter, the ladder 
approach, which accounts for the reduced magnetic frustration, yields 
the $s_{\pm}$ order parameter in agreement with the experimental data.

As a final remark, we note that consideration of SU(2) 
symmetric interaction and non-local self-energy \cite{SCDGA} 
(or $\lambda$ correction) effects \cite{Lcorr} in the ladder approximation, as well as the frequency 
dependence of superconducting gap,  would allow to obtain realistic
superconducting transition temperatures. The extension of the theoretical 
approach used in the present study to pnictides and chalcogenides 
with stronger electronic correlations than in LaFeAsO would also 
require accounting for the damping of electronic excitations in the 
Bethe-Salpeter equation, which becomes progressively stronger with 
increasing degrees of local magnetic moment formation. Apart from that, 
the consideration of 
the effect of doping is interesting.
This constitutes the subject of future studies. 
\\
\appendix
\renewcommand\theequation{A\arabic{equation}}
\setcounter{equation}{0}

\section*{Appendix. Density-density form of the second-order pairing vertex}
In the second-order perturbation theory we obtain the following expressions for the second-order pairing vertices in singlet and triplet channels
\begin{align}
V_{\mathbf{k,p},\alpha \beta }^{s} &=\frac{1}{2}\sum\limits_{mm^{\prime
},s=\pm}\left( 2V_{\mathbf{k},-\mathbf{k};s\mathbf{p},-s\mathbf{p,\uparrow
\downarrow }}^{mm^{\prime }}-V_{\mathbf{k},-\mathbf{k};s\mathbf{p},-s\mathbf{%
p,\uparrow \uparrow }}^{mm^{\prime }}\right)  \notag \\
&\times \left( S_{\mathbf{k}}^{m\alpha }\right) ^{\ast }(S_{-\mathbf{k}%
}^{m^{\prime }\alpha })^{\ast }S_{\mathbf{p}}^{m\beta }S_{-\mathbf{p}%
}^{m^{\prime }\beta }. \\
V_{\mathbf{k,p},\alpha \beta }^{t} &=\frac{1}{2}\sum\limits_{mm^{\prime
}}\left( V_{\mathbf{k},-\mathbf{k};\mathbf{p},-\mathbf{p,\uparrow \uparrow }%
}^{mm^{\prime }}-V_{\mathbf{k},-\mathbf{k};-\mathbf{p},\mathbf{p,\uparrow
\uparrow }}^{mm^{\prime }}\right)\notag \\
&\times \left( S_{\mathbf{k}}^{m\alpha }\right)
^{\ast }(S_{-\mathbf{k}}^{m^{\prime }\alpha })^{\ast }S_{\mathbf{p}}^{m\beta
}S_{-\mathbf{p}}^{m^{\prime }\beta }.
\end{align}%
where the matrices $S^{m\alpha}_{\mathbf k}$ diagonalize the Hamiltonian (see main text),
\begin{equation}
V_{2,\mathbf{k},-\mathbf{k};\mathbf{p},-\mathbf{p,\sigma \sigma }^{\prime
}}^{mm^{\prime }}=-\sum_{\widetilde{m}\widetilde{m}^{\prime },\sigma
^{\prime \prime }}V_{0,\sigma \sigma ^{\prime \prime }}^{m\widetilde{m}%
}L_{ph}^{\widetilde{m}\widetilde{m}^{\prime }}(\mathbf{k}-\mathbf{p}%
)V_{0,\sigma ^{\prime \prime }\sigma ^{\prime }}^{\widetilde{m}^{\prime
}m^{\prime }},
\end{equation}%
is the respective 2-particle vertex,
\begin{align}
L_{ph}^{mm^{\prime }}(\mathbf{q})&=\sum_{i\nu_n}\chi^{0,mm'}_{\nu,{\mathbf q}}, 
\end{align}%
is the bubble ($\chi^{0,mm'}_{\nu,{\mathbf q}}$ is defined in the main text), and
\begin{equation}
V_{0,\sigma \sigma ^{\prime }}^{mm^{\prime }}=U_{\uparrow\downarrow}^{mm^{\prime }}\delta
_{\sigma ,-\sigma ^{\prime }}+U_{\uparrow\uparrow}^{mm^{\prime }}\left(
1-\delta _{mm^{\prime }}\right) \delta _{\sigma \sigma ^{\prime }}
\end{equation}%
is the bare interaction.

\vspace{1cm}
\section*{ACKNOWLEDGMENTS}
The DFT+DMFT calculations of the spectral properties and spin susceptibility 
were carried out within the framework of the state assignment of the Ministry 
of Science and Higher Education of the Russian Federation for the IMP UB RAS. 
Calculations of the tendency to superconducting pairing and symmetry of the superconducting 
gap were supported by the Russian Science Foundation (Project 24-12-00024).

\end{document}